\documentclass[11pt]{article}
\usepackage{amsmath}
\usepackage{amssymb}
\usepackage{bbm}
\usepackage{epsfig}
\usepackage{graphicx}
\allowdisplaybreaks[4]
\begin{document}
\title{Some remarks on tree-level vacuum stability in two Higgs
doublet models${}^{1}$} \footnotetext[1]{Prepared for the
Proceedings of The International Conference on High Energy and
Mathematical Physics, Marrakech, Morocco, 3-7 April 2005. Talk
presented by R. Santos.}
\author{A. Barroso~\footnote{barroso@cii.fc.ul.pt},
P.M. Ferreira~\footnote{ferreira@cii.fc.ul.pt} and R.
Santos~\footnote{rsantos@cii.fc.ul.pt}
\\
Centro de F\'{\i}sica Te\'orica e Computacional, \\
Universidade de Lisboa, Av. Prof. Gama Pinto, 2, 1649-003 Lisboa, Portugal \\
}
\date{June, 2005}
\maketitle \noindent {\bf Abstract.} It is proved that the minimum
of a general two Higgs doublet models' potential is stable at tree
level. A relation between stability and flavour changing neutral
currents at tree level is shown.

\section{Introduction}

One of the most straightforward ways to extend the Standard Model
of the weak interactions (SM) is to add a second Higgs doublet to
the scalar sector. This type of models is known in the literature
as two Higgs doublet models (2HDM). They present a richer
phenomenology due to the appearance of charged and also
pseudo-scalar Higgs. However, maybe one of the main reasons of
interest in this class of models is the possibility of having
spontaneous CP violation~\cite{lee}, thus helping to solve the
baryogenesis problem~\cite{err} (for a review, see~\cite{sher}).
These models have a very large number of independent parameters.
The most general 2HDM has 14 real parameters although, with a
particular choice of basis, one can reduce this number to 11
independent parameters (see, for instance,~\cite{habdav}). There
are some bounds on the parameters in models derived from the
general one by imposing a $Z_2$ or a $U(1)$ symmetry. However,
besides some very weak experimental and theoretical bounds, very
little is known about their allowed values, especially in this
most general case. The same is true for Supersymmetric models,
where the parameter space is generally larger. One idea that has
been applied to Supersymmetric theories to restrict their allowed
parameter space is to use charge and colour breaking (CCB) bounds.
If a given combination of parameters causes the appearance in the
potential of a minimum where charged/coloured fields have vacuum
expectation values (vevs), then that combination should be
rejected. This appealing idea was introduced by Fr\'ere {\em et
al}~\cite{fre} and applied, in numerous papers, to several
supersymmetric theories~\cite{eve}. Phenomenological analysis of
supersymmetric Higgs masses use this tool to increase the models'
predictive power~\cite{phe}. It is therefore of interest to apply
similar techniques to the 2HDM and try to limit its parameter
space. The scalars of this theory have no colour quantum numbers
but there are charged fields so charge breaking (CB) extrema are
in principle possible. It was shown in ref.~\cite{vel} that to
assure that there were no stationary points corresponding to
charge or CP spontaneous breaking, one had to restrict the
parameter space to 7 independent parameters. This led to two
independent potentials, stable under renormalisation, as they were
protected by a $Z_2$ or a global $U(1)$ symmetries (that can be
softly broken). It is interesting to stress that these are the
usual symmetries introduced to prevent flavour changing neutral
currents. We will come back to this point later.

\section{Tree-level stability}

This section follows very closely the work done in ~\cite{pot10,
pot14}. There are two scalar Higgs doublets in the theory,
$\Phi_1$ and $\Phi_2$, both having hypercharge $Y=1$, i.e., 
\begin{equation}
\Phi_1 = \begin{pmatrix} \varphi_1 + i \varphi_2 \\ \varphi_5 + i
\varphi_7
\end{pmatrix} \;\; , \;\; \Phi_2 = \begin{pmatrix} \varphi_3 + i \varphi_4 \\
\varphi_6 + i \varphi_8 \end{pmatrix} \;\; .
\end{equation}
The numbering of the real scalar $\varphi$ fields is chosen for
convenience in writing the mass matrices for the scalar particles.
Our basis is obtained by first writing down the four $SU(2)_W
\times U(1)_Y$ invariants one can construct from these two
doublets, namely $x_1\,\equiv\,|\Phi_1|^2$,
$x_2\,\equiv\,|\Phi_2|^2$, $x_3\,\equiv\,Re(\Phi_1^\dagger\Phi_2)$
and $x_4 \,\equiv\,Im(\Phi_1^\dagger\Phi_2)$. Notice that under a
CP transformation ($\Phi_1 \rightarrow \Phi_1^*\;,\;\Phi_2
\rightarrow \Phi_2^*$) the invariants $x_1$, $x_2$ and $x_3$
remain the same but $x_4$ changes sign. It is now a simple matter
to write down the most general 2HDM model, i.e., 
\begin{align}
V \;\;=& \;\; a_1\, x_1\, + \,a_2\, x_2\, + \,a_3 x_3 \,+\, a_4
x_4 \,+\, b_{11} \, x_1^2\, +\, b_{22}\, x_2^2\, +\, b_{33}\,
x_3^2\, +\, b_{44}\, x_4^2\, +\,
\nonumber \\
 & \;\; b_{12}\, x_1 x_2\, +\, b_{13}\, x_1 x_3\, + b_{14}\,
x_1 x_4\, +\,b_{23}\, x_2 x_3 +\,b_{24}\, x_2 x_4 +\,b_{34}\, x_3
x_4\;\; , \label{eq:pot}
\end{align}
where the $a_i$ parameters have dimensions of mass squared and the
$b_{ij}$ parameters are dimensionless. The terms linear in $x_4$
are those that break CP explicitly, and eliminating them we are
left with the CP preserving potential with 10 parameters that we
have used in reference~\cite{pot10}. Notice that this potential
has only 9 independent parameters due to basis invariance (see
\cite{habdav} for details). For convenience we introduce a new
notation, namely a vector A and a square matrix B, given by 
\begin{equation}
A\;=\;\begin{bmatrix} a_1 \\ a_2 \\a_3 \\ a_4 \end{bmatrix} \;\;\;
, \;\;\; B\;=\;\begin{bmatrix} 2 b_{11} & b_{12} & b_{13} & b_{14}
\\ b_{12} & 2 b_{22} & b_{23} & b_{24} \\ b_{13} & b_{23} & 2
b_{33} & b_{34} \\ b_{14} &  b_{24} & b_{34} & 2 b_{44}
\end{bmatrix} \;\; .
\label{eq:ab}
\end{equation}
Defining the vector $X\, =\, (x_1\,,\,x_2\,,\, x_3\,,\,x_4)$, we
can rewrite the potential~\eqref{eq:pot} as 
\begin{equation}
V \;=\; A^T\,X \;+\; \frac{1}{2}\,X^T \,B\,X \;\;\; .
\label{eq:vm}
\end{equation}
It is a well known fact that the 2HDM potential has only three
types of possible minima~\cite{lee,sher}. With our conventions we
can define a charge breaking (CB) minimum configuration as
\begin{equation}
\Phi_1 = \begin{pmatrix} 0 \\ v^{'}_1
\end{pmatrix} \;\; , \;\; \Phi_2 = \begin{pmatrix} \alpha \\
v^{'}_2 \end{pmatrix} \;\; .
\end{equation}
The vev $\alpha$ breaks the $U(1)_{em}$ symmetry and so gives a
mass to the photon. In the second type of minima only neutral
fields have vevs, and there are two different possibilities which
we define as
\begin{equation}
\Phi_1 = \begin{pmatrix} 0 \\ v_1
\end{pmatrix} \;\; , \;\; \Phi_2 = \begin{pmatrix} 0 \\
v_2  \end{pmatrix} ,
\end{equation}
and
\begin{equation}
\Phi_1 = \begin{pmatrix} 0 \\ v^{''}_1+ i \, \delta
\end{pmatrix} \;\; , \;\; \Phi_2 = \begin{pmatrix} 0 \\
v^{''}_2 \end{pmatrix} \;\; .
\end{equation}
We call the first the $N_1$ minimum, and the second the $N_2$
minimum. When the model is reduced to the explicit CP preserving
potential, $N_1$ is the CP preserving minimum and $N_2$ is the one
that spontaneously breaks CP. For the potential where CP is
explicitly broken from the start, there is no physical distinction
between the $N_1$ and $N_2$ minima.

The demonstration that if one of the normal minima exists, it is
certainly deeper than the charge breaking one was done
in~\cite{pot10, pot14}. We will skip here some non-essential
intermediate steps and refer the reader to these articles. Let
$V^\prime$ be a vector with components $V^\prime_i =
\partial V/
\partial x_i$, evaluated at the $N_1$ minimum. At $N_1$ the
non-zero vevs are $\varphi_5 = v_1$ and $\varphi_6 = v_2$, so that
$x_1 = v_1^2$, $x_2 = v_2^2$, $x_3 = v_1 v_2$ and $x_4 = 0$. We
define the vector $X_{N_1}$ as the value of $X$ at this minimum,
and it is a trivial matter to demonstrate that the value of the
potential at the minimum is given by 
\begin{equation}
V_{N_1} \;\; = \;\; \frac{1}{2}\,A^T\,X_{N_1} \;\; = \;\;
-\frac{1}{2}\, X^T_{N_1}\, B\,X_{N_1} \;\;. \label{eq:vmin}
\end{equation}
Further, we can write down the non-trivial stationarity
conditions, which are 
\begin{equation}
\begin{array}{rclcl}
\displaystyle{\frac{\partial V}{\partial v_1}}= 0 &\Leftrightarrow
& V^\prime_1 \displaystyle{\frac{\partial x_1}{\partial v_1}}
\,+\, V^\prime_3 \displaystyle{ \frac{\partial x_3}{\partial v_1}}
= 0 &\Leftrightarrow &  V^\prime_1  =
\left( \displaystyle{-\frac{V^\prime_3}{2 v_1 v_2}} \right)v_2^2\vspace{0.2cm}\\
\displaystyle{\frac{\partial V}{\partial v_2}} = 0
&\Leftrightarrow & V^\prime_2 \displaystyle{\frac{\partial
x_2}{\partial v_2}} \,+\,V^\prime_3 \displaystyle{\frac{\partial
x_3}{\partial v_2}} = 0 &\Leftrightarrow & V^\prime_2  =
\left(\displaystyle{-\frac{V^\prime_3}{2 v_1 v_2}}\right) v_1^2
\vspace{0.2cm}\\
\displaystyle{\frac{\partial V}{\partial \varphi_7}} = 0
&\Leftrightarrow & V^\prime_4 \displaystyle{\frac{\partial
x_4}{\partial \varphi_7}} = 0 &\Leftrightarrow &  V^\prime_4 = 0
\;\;\;\;\;\; . \label{eq:min}
\end{array}
\end{equation}
From eq.~\eqref{eq:vmin} we see that $V^\prime \,=\, A\,+\,B\,
X_{N_1}$ and from the equations above we can obtain 
\begin{equation}
V^\prime \; = \; \begin{bmatrix} V^\prime_1 \\ V^\prime_2 \\ V^\prime_3 \\
V^\prime_4 \end{bmatrix} \; = \; -\frac{V^\prime_3}{2 v_1 v_2}\,
\begin{bmatrix} v_2^2 \\ v_1^2 \\ - 2 v_1 v_2 \\ 0 \end{bmatrix}
\;\; . \label{eq:vl}
\end{equation}
Written in this form we see, plainly, that $V^\prime_1$ and
$V^\prime_2$ have the same sign. Now, in reference~\cite{pot10} we
have obtained general expressions for the mass matrices of the
theory's scalar particles. In particular we have shown that
$M^2_{H^\pm}\, =\, V^\prime_1\,+\,V^\prime_2\, =\, -V^\prime_3
v^2/(2 v_1 v_2)$, with $v^2\,=\,v_1^2\,+\, v_2^2$. If $N_1$ is a
minimum then all of the squared scalar masses are positive and so
this quantity is positive. Another consequence of the minimisation
conditions is that we obtain $X_{N_1}^T\, V^\prime \,=\,0$.

Regarding the CB stationary point, the fields that have non-zero
vevs are now $\varphi_5 = v^\prime_1$, $\varphi_6 = v^\prime_2$
and $\varphi_3 = \alpha$. We define the vector $Y$ to be equal to
the vector $X$ evaluated at this stationary point, that is, $Y$
has components $Y = ({v^\prime_1}^2 \,,\,{v^\prime_2}^2\,+\,
\alpha^2 \,,\,v^\prime_1 v^\prime_2 \,,\, 0)$. The stationarity
conditions now give 
\begin{equation}
\begin{array}{rclcl}
\displaystyle{\frac{\partial V}{\partial v^\prime_1}}= 0
&\Leftrightarrow & V^\prime_1 \displaystyle{\frac{\partial
x_1}{\partial v^\prime_1}} \,+\, V^\prime_3
\displaystyle{\frac{\partial x_3}{\partial v^\prime_1}} = 0
&\Leftrightarrow &  V^\prime_1  = \left(
\displaystyle{-\frac{V^\prime_3}{2
v^\prime_1 v^\prime_2}} \right) {v^\prime_2}^2 \vspace{0.2cm} \\
\displaystyle{\frac{\partial V}{\partial v^\prime_2}} = 0
&\Leftrightarrow & V^\prime_2 \displaystyle{\frac{\partial
x_2}{\partial v^\prime_2}} \,+\, V^\prime_3
\displaystyle{\frac{\partial x_3}{\partial v^\prime_2}} = 0
&\Leftrightarrow & V^\prime_2  =
\left(\displaystyle{-\frac{V^\prime_3}{2
v^\prime_1 v^\prime_2}}\right) {v^\prime_1}^2 \vspace{0.2cm} \\
\displaystyle{\frac{\partial V}{\partial \alpha}} = 0
&\Leftrightarrow & V^\prime_2 \displaystyle{\frac{\partial
x_2}{\partial \alpha}} = 0
&\Leftrightarrow & V^\prime_2 = 0 \vspace{0.2cm} \\
\displaystyle{\frac{\partial V}{\partial \varphi_7}} = 0
&\Leftrightarrow & V^\prime_4 \displaystyle{\frac{\partial
x_4}{\partial \varphi_7}} = 0 &\Leftrightarrow &  V^\prime_4 = 0
\;\;\;\;\;\; .
\end{array}
\end{equation}
We thus obtain, for the CB stationary point, $V^\prime_i = 0$. The
equation that determines $Y$ is simply $A\; + \; B\,Y \;\; = \;\;
0$, which implies that, even for this more complex potential, the
CB stationary point, if it exists, is unique. The value of the
potential at this charge breaking stationary point is given by 
\begin{equation}
V_{CB} \;\; = \;\; \frac{1}{2}\,A^T\,Y \;\; = \;\;
-\frac{1}{2}\,Y^T\, B\,Y \;\;. \label{eq:vcb}
\end{equation}
Remembering that $X_{N_1}^T\,V^\prime \,=\,0$ we obtain, from
$V^\prime \,=\, A\,+\,B\, X_{N_1}$ and $A\; + \; B\,Y \;\; = \;\;
0$, that 
\begin{equation}
X_{N_1}^T\,B\, Y \;\; = \;\; X_{N_1}^T\,B\, X_{N_1} \;\; = \;\;
-\,2\,V_{N_1} \;\; . \label{eq:vn1}
\end{equation}
We can calculate the quantity $Y^T\,V^\prime$, which is easily
seen to be given by 
\begin{equation}
Y^T\,V^\prime \;\; = \;\; -\,Y^T\,B\, Y  \,+\, Y^T\,B\, X_{N_1}
\;\;\; .
\end{equation}
But, from eq.~\eqref{eq:vcb}, it follows that $Y^T\,B\,Y \,=\,
-\,2\,V_{CB}$ and eq.~\eqref{eq:vn1} and the fact that the matrix
B is symmetric gives $Y^T\,B\, X_{N_1}\,=\, -\,2\,V_{N_1}$.
Therefore, we reach the conclusion that 
\begin{equation}
V_{CB}\;-\;V_{N_1} \;\; = \;\;\frac{1}{2}\,Y^T\,V^\prime \;\; =
\;\; \frac{M^2_{H^\pm}}{2\,v^2} \;\left[ (v^\prime_1\,v_2
\;-\;v^\prime_2\,v_1)^2\; + \; \alpha^2\,v_1^2\right]\;\;\; .
\label{eq:difp}
\end{equation}
Then, it is clear that, if $N_1$ is a minimum of the theory, all
of its squared masses will be positive, and therefore we will have
$V_{CB}\;-\;V_{N_1} \;> \;0$, which implies that the charge
breaking stationary point, when it exists, is always located above
the $N_1$ minimum. Furthermore, it is easy to obtain the equality
$Y\,=\,X\,-\, B^{-1}V^\prime$, so that $Y^T\,V^\prime$ becomes
equal to $-{V^\prime}^T\,B^{-1} \,V^\prime$. In ref.~\cite{pot10}
we demonstrated that the matrix $B$ determines the nature of the
CB stationary point. The equality we have just obtained
demonstrates that the matrix $B$ is {\em not} positive definite.
For reasons explained in~\cite{pot10} it cannot be negative
definite (which arises from requiring that the potential we are
working with is bounded from below), which implies that $B$ is
neither positive nor negative definite. As a result, the CB
stationary point is a saddle point.

Now we turn our attention to the $N_2$ minimum. {\em A priori}
there is no reason why the 2HDM potential cannot have,
simultaneously, both ``normal" minima, so the question arises, can
the potential be in an $N_2$ minimum that is not deeper than a CB
stationary point? The answer is no, and the demonstration follows
very closely the one we just concluded. For the $N_2$ minimum, the
fields that have non-zero vevs are $\varphi_5 = v^{\prime
\prime}_1$, $\varphi_6 = v^{\prime \prime}_2$ and  $\varphi_7 =
\delta$, so that the $X$ vector is now given by $X_{N_2}\,=\,
({v^{\prime \prime}_1}^2\,+\,\delta^2\,,\,{v^{\prime
\prime}_2}^2\,,\,v^{\prime \prime}_1\, v^{\prime \prime}_2\,,\,
-\,v^{\prime \prime}_2\,\delta)$. Solving the stationarity
conditions as before, we find that the vector $V^\prime\,=\,
A\,+\,B\, X_{N_2}$, at this minimum, is given by 
\begin{equation}
V^\prime_{N_2} \; = \; \begin{bmatrix} V^\prime_1 \\ V^\prime_2 \\ V^\prime_3 \\
V^\prime_4 \end{bmatrix} \; = \;
-\frac{\left(V^\prime_3\right)_{N_2}}{2 v^{\prime \prime}_1
v^{\prime \prime}_2} \,
\begin{bmatrix} x_2 \\ x_1 \\ - 2\, x_3 \\ -2\, x_4 \end{bmatrix}
\;\; ,
\end{equation}
and in fact this final expression also applies to the vector
$V^\prime$, evaluated at the $N_1$ minimum. We still have
$X_{N_2}^T\,V^\prime_{N_2}\,=\,0$ and $-\left(V^\prime_3
\right)_{N_2}/(2 v^{\prime \prime}_1 v^{\prime
\prime}_2)\;=\;\left(M^2_{H^\pm}/v^2\right)_{N_2}$. In this
expression the charged scalar mass is the non-zero eigenvalue of
the charged mass matrix at the $N_2$ minimum and $v^2$ is now
given by $v^2\,=\, {v^{\prime \prime}}^2_1\,+\,{v^{\prime
\prime}}^2_2\,+\,\delta^2$. We are therefore in the exact
conditions of the $N_1$ minimum and conclude, likewise, that 
\begin{equation}
V_{CB}\;-\;V_{N_2} \;\; = \;\;\frac{1}{2}\,Y^T\,V^\prime \;\; =
\;\; \left(\frac{M^2_{H^\pm}}{2\,v^2}\right)_{N_2} \;\left[
(v^{\prime \prime}_1\,v_2 \;-\;v^{\prime \prime}_2\,v_1)^2\; + \;
\alpha^2\,(v_1^2\;+\;\delta^2)\;+\;\delta^2\, v^{\prime \prime \,
2}_2\right]\;\;>\;\;0 . \label{eq:difn2}
\end{equation}
Again, the charge breaking stationary point lies above the normal
minimum, and again it is a saddle point, for the same reasons we
have explained before. Unfortunately we cannot apply this
procedure to determine whether one of the minima $N_1$ or $N_2$ is
deeper than the other. If one follows the steps we have outlined,
one is left with
\begin{equation}
V_{N_2}\;-\;V_{N_1} \;\; = \;\;\frac{1}{2}\left[
\left(\frac{M^2_{H^\pm}}{v^2}\right)_{N_1}\;-\;\left(\frac{M^2_{H^\pm}}{v^2}
\right)_{N_2} \right] \;\left[ (v^{\prime \prime}_1\,v_2
\;-\;v^{\prime \prime}_2\,v_1)^2\; + \;
\delta^2\,v_2^2\right]\;\;\; . \label{eq:difcp}
\end{equation}
If both $N_1$ and $N_2$ are minima - and there does not seem to be
anything preventing it - the terms proportional to $M^2_{H^\pm}$
will be positive, but it seems impossible to tell which one is the
largest.

For the special case of a potential without explicit CP breaking,
the $N_2$ stationary point is the one with spontaneous CP
breaking. The $N_1$ stationary point preserves both charge and CP
and it is what we called, in ref.~\cite{pot10}, the normal
minimum. In that reference we calculated the mass matrices for the
several minima possible and showed that
$(M^2_{H^\pm}/v^2)_{N_2}\;=\;-\,b_{44}$. At the normal minimum we
have $M^2_A\;=\;M^2_{H^\pm}\,+\,b_{44}\,v^2$, $M^2_A$ being the
squared mass of the pseudoscalar. Then, in this case,
eq.~\eqref{eq:difcp} gives the difference of the values of the
potential at the spontaneous CP breaking stationary point and at
the normal one, and reduces to 
\begin{equation}
V_{CP}\;-\;V_{N} \;\; = \;\;\frac{M^2_A}{2\,v^2}\;\left[
(v^{\prime \prime}_1\,v_2 \;-\;v^{\prime \prime}_2\,v_1)^2\; + \;
\delta^2\,v_2^2\right]\;\;\; . \label{eq:ma}
\end{equation}
It is clear that if there exists a normal minimum, $M^2_A$ is
positive and the CP stationary point is above the normal minimum.
It was also shown in ref.~\cite{pot10} that the CP stationary
point is necessarily a saddle point, analogously to what happens
with the CB case.

\section{FCNC and stability}

In this section we want to stress a result obtained in~\cite{vel}:
flavour conservation can be achieved by demanding only natural CP
conservation and no CB in the absence of fermions, that is, the
potentials possessing only CP and CB invariant minima are
consistent with the absence of flavour changing neutral currents
(FCNC) in the tree level Yukawa couplings. The problem we are
addressing is how to force the CB and CP stationarity conditions
to have no solution regardless of what the values of the
parameters may be. The number of ways to accomplish it is
obviously huge. However, the conditions chosen will be useless if
they are not preserved by renormalization. Thus, the only safe way
to do it is by means of imposing some kind of symmetry to the
potential. The interesting point is that it is possible to enforce
the stationary point conditions to have no solution by demanding
invariance of the tree-level potential to a $Z_2$ or to a global
$U(1)$ symmetry. That way, the stationary point conditions for the
CP and CB cases have no solution at any order in perturbation
theory. We have proved in the previous section that a normal
minimum is stable in a more general model where CP can be
spontaneously broken. However, in that more general model, there
is also room for a stable CP minimum. By imposing the $Z_2$ or a
global $U(1)$ symmetry to the potential written in our basis this
possibility ceases to exist. The only allowed minimum is the
charge and CP conserving one.

The most general Yukawa lagrangian of a general 2HDM gives rise to
the appearance of FCNC which are known to be severely constrained
by experiment. However, the same symmetries that were used to
prevent the existence of the CB and CP stationary points, can now
be used to avoid in a natural way the appearance of FCNC at tree
level. This can be done by imposing similar symmetries to the
appropriate fermion fields.

This way we build two different models which we call A and
B~\footnote{In ref. \cite{vel} we have named those models I and
II. However, this notation was misleading since the same numbering
is used for the different Yukawa lagrangians of 2HDM.} in
~\cite{rui}, where model A is based on a $Z_2$ symmetry and model
B is based on a global $U(1)$ symmetry, softly broken by a
dimension two term (otherwise an axion would be produced). They
both have 7 independent parameters, are tree-level stable and
renormalizable. There are no differences between the two models in
the gauge and Yukawa sectors. However, they present quite a
different set of Feynman rules in the scalar sector. For instance,
the difference between the strength of the vertex $h H^+ H^-$ in
the two models is
\begin{equation}
(g_{h H^+ H^-})_A - (g_{h H^+ H^-})_B \, = \, 2i \,
\frac{M^2_A}{v^2} \,  \frac{\cos(\alpha + \beta)}{\sin (2 \beta)}
\end{equation}
where $h$ is the lightest CP-even scalar.

To finish this section, we would like to point out that whereas in
model A the requirement of boundness from below is automatically
fulfilled, in model B the same requirement needs a condition
between the masses that reads
\begin{equation}
M^2_h + M^2_H \geq M^2_A  \enskip
\end{equation}
where $H$ is the heaviest CP-even Higgs.

\section{Conclusions}

\begin{figure}[htb]
  \begin{center}
    \epsfig{file=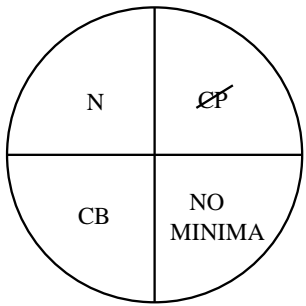,width=4 cm,angle=0}
    \caption{The quiet world of Two-Higgs doublet models.}
    \label{fig_new1}
  \end{center}
\end{figure}

In fig.1 we present the result of our work for the potential that
does not break CP explicitly. We have shown that the four
different worlds do not intersect each other. Each region
corresponds to a theory perfectly stable at tree level. Once the
world has chosen to be in one of those minima it will remain there
unless the values of the parameters of the potential change. If we
are for instance in the normal minimum then the model is protected
against electric charge or CP spontaneous breaking. In other
words, if the model has a minimum preserving $U(1)_{em}$ and CP,
that minimum is global. In this way, there is absolutely no
possibility of tunnelling to deeper minima, and, for instance, the
masslessness of the photon is guaranteed in these models. Notice
however that the same is true for the CP or the CB minimum. If a
CB or a CP minimum exists, it is now the global one and it is
perfectly stable. Again, no tunnelling occurs.

Charge breaking would be disastrous but there is considerable
interest, from cosmologists to particle physicists, in models with
the possibility of spontaneous CP violation. We have determined
that this cannot happen for those ranges of parameters that lead
to Normal minima. However, we have also established a very precise
condition for spontaneous CP breaking to occur: CP is
spontaneously broken if and only
${V^\prime}^T\,B_{CP}^{-1}\,V^\prime > 0$~\footnote{In
reference~\cite{sil} CP violating quantities involving only the
Higgs sector were derived in models with explicit CP violation.}.
In these circumstances the 2HDM no longer has a normal minimum,
nor a charge breaking one, as we have shown.

In the most general potential there are just the normal minima (CP
violating ones) and the CB minimum. There we have proved that the
same result holds. It is interesting to point out the following
aspect of these results. If one observes equations~\eqref{eq:difp}
and~\eqref{eq:difn2}, one sees that the difference in the depth of
the potential between the normal minimum and the CB stationary
point is controlled by the charged Higgs squared mass. On the
other hand, equation~\eqref{eq:ma} shows that the potential depth
difference between the CP and the normal stationary points is
proportional to the pseudoscalar squared mass. That is, the depth
of the potential at a stationary point that breaks a given
symmetry, relative to the normal minimum, depends, in a very
straightforward way, on the mass of the scalar particle directly
linked with that symmetry.

To finish, let us also stress that our conclusions are absolutely
general, independent of particular values of the parameters of the
theory, obviously. They hold for any of the more restricted models
considered in ref.~\cite{vel}. It is simple to recover the
conditions presented in that reference to avoid CP minima by
analysing the matrix $B_{CP}$. We remark that the Higgs potential
of the Supersymmetric Standard Model (SSM) is also included in the
potentials we studied - in fact, it corresponds to the case
$b_{11}\,=\, b_{22}\, =\,-\, b_{12}/2\,=\,M_Z^2/(2 v^2)$,
$b_{33}\,=\,b_{44}\,=\,2\,M_W^2/v^2$ and the remaining $b$
parameters set to zero, following the conventions of
ref.~\cite{cast}. So we could conclude that at tree-level, the
Supersymmetric Higgs potential is safe against charge of CP
violation, though this would not preclude charge, colour or CP
breaking arising from other scalar fields present in those models.
However, we must be cautious: it has been shown~\cite{gam} that
one-loop contributions to the minimisation of the potential have
an enormous impact on charge breaking bounds in Supersymmetric
models. Also, it was recently shown~\cite{eu} that unless one
performs a full one-loop calculation (both for the potential and
the vevs, in both the CB potential and the ``normal" one) the
bounds one obtains can be overestimated. Therefore, we urge
caution in applying these conclusions to the SSM. Nevertheless one
would expect the one-loop contributions to be much less important
in the non-supersymmetric 2HDM due to the much smaller particle
content of the latter theory.

\vspace{0.25cm} {\bf Acknowledgments:} We are thankful to Pedro
Freitas, Lu\'{\i}s Trabucho and Jo\~ao Paulo Silva for their
assistance and discussions. This work is supported by
Funda\c{c}\~ao para a Ci\^encia e Tecnologia under contracts
PDCT/FP/FNU/50155/2003 and POCI/FIS/59741/2004. P.M.F. is
supported by FCT under contract SFRH/BPD/5575/2001.

\end{document}